\documentclass[aip,apl,twocolumn,superscriptaddress]{revtex4}
\usepackage{graphicx}
\usepackage{morefloats}
\usepackage{color}
\usepackage{amssymb}
\usepackage{float}

\begin{document}


\title{Magnetic phase transitions in Ta/CoFeB/MgO multilayers}


\author{I.\,Barsukov}
\affiliation{Physics and Astronomy, University of California, Irvine, CA 92697, USA}
\author{Yu\,Fu}
\affiliation{INAC/CEA, Grenoble, 17 avenue des Martyrs, 38054 Grenoble, France}
\author{C.\,Safranski}
\affiliation{Physics and Astronomy, University of California, Irvine, CA 92697, USA}
\author{Y.-J.\,Chen}
\affiliation{Physics and Astronomy, University of California, Irvine, CA 92697, USA}
\author{B.\,Youngblood}
\affiliation{Physics and Astronomy, University of California, Irvine, CA 92697, USA}
\author{A.\,M.\,Gon\c{c}alves}
\affiliation{Physics and Astronomy, University of California, Irvine, CA 92697, USA}
\affiliation{Centro Brasileiro de Pesquisas F\'{i}sicas, Rua Dr.\,Xavier Sigaud, 150, Rio de Janeiro 22.290-180, RJ, Brazil}
\author{M.\,Spasova}
\affiliation{Fakult\"{a}t f\"{u}r Physik and Center for Nanointegration (CeNIDE), Universit\"at Duisburg-Essen, 47048 Duisburg, Germany}
\author{M.\,Farle}
\affiliation{Fakult\"{a}t f\"{u}r Physik and Center for Nanointegration (CeNIDE), Universit\"at Duisburg-Essen, 47048 Duisburg, Germany}
\author{J. A. Katine}
\affiliation{HGST Research Center, San Jose, California 95135, USA}
\author{C.\,C.\,Kuo}
\affiliation{Components Research, Intel, Hillsboro, OR  97124, USA}
\author{I.\,N.\,Krivorotov}
\affiliation{Physics and Astronomy, University of California, Irvine, CA 92697, USA}



\begin{abstract}
We study thin films and magnetic tunnel junction nanopillars based on Ta/Co$_{20}$Fe$_{60}$B$_{20}$/MgO multilayers by electrical transport and magnetometry measurements. These measurements suggest that an ultrathin magnetic oxide layer forms at the Co$_{20}$Fe$_{60}$B$_{20}$/MgO interface. At approximately 160\,K, the oxide undergoes a phase transition from an insulating antiferromagnet at low temperatures to a conductive weak ferromagnet at high temperatures. This  interfacial magnetic oxide is expected to have significant impact on the magnetic properties of CoFeB-based multilayers used in spin torque memories.
\end{abstract}



\maketitle

The ferromagnet/oxide (FM/Ox) interfaces play an important role in modern spintronics. Insulating oxide layers sandwiched between two metallic ferromagnets form magnetic tunnel junctions (MTJs)\cite{Parkin2004, Yuasa,CuiRalph} which find applications in magnetic sensors \cite{Prokopenko2012,Yuasa2004,Miwa2014, Zhang2011}, spin torque oscillators \cite{Deac,Nazarov,Zeng,Rowlands}, and non-volatile spin torque memory (STT-RAM) \cite{ApalkovSTTMRAM}. A number of FM/Ox interfaces exhibit large perpendicular magnetic anisotropy (PMA)  \cite{Ikeda2010,Zhu2012,Ikeda2008, Worledge} needed for enhancing thermal stability of STT-RAM devices  \cite{Mangin2006, Heindl, ZhaoWang2011}. Furthermore, some FM/Ox interfaces exhibit magneto-electric coupling that allows control of the interfacial PMA with an electric field applied perpendicular to the interface. This effect, known as voltage-controlled magnetic anisotropy (VCMA) \cite{DuanTsymbal, Endo2010, Nozaki, Maruyama}, can be used for energy-efficient voltage-driven switching of magnetization at low current densities \cite{Zhu2012, Wang2012,Shiota}.

The interface between Co$_x$Fe$_y$B$_z$ (CoFeB) ferromagnet and MgO insulator is one of the most important interfaces in spintronics because the STT-RAM technology is based on CoFeB/MgO/CoFeB MTJs \cite{Yakushiji2010}. Comprehensive understanding of structural, magnetic and electronic properties of CoFeB/MgO-based multilayers is at the forefront of applied spintronics research. In this Letter, we report magnetometry and electrical transport studies of (i) CoFeB/MgO/CoFeB nanoscale magnetic tunnel junctions and (ii) CoFeB/MgO based multilayer films. These studies reveal surprising anomalies in the temperature dependence of resistance and magnetization of the CoFeB films interfaced with MgO. Our data suggest that ultrathin magnetic oxide layers are formed at the CoFeB/MgO interfaces prepared under typical deposition and annealing conditions employed in fabrication of CoFeB/MgO/CoFeB MTJs with high tunneling magnetoresistance (TMR) \cite{Parkin2004, Yuasa}. Interestingly, the high TMR observed in these MTJs seems to be largely unaffected by the interfacial oxide formation. 

We study two CoFeB/MgO-based systems: nanoscale CoFeB/MgO/CoFeB MTJs and CoFeB/MgO based multilayer films. The nanoscale 60$\times$170\,nm$^2$ elliptical MTJs are patterned by ion milling from a Ta(5)/ SAF/ MgO(0.82)/ FL/ Ta(5) multilayer \cite{multilayer} (thicknesses in nm) deposited on Si/ SiOx by magnetron sputtering in a Singulus TIMARIS system. Here SAF $\equiv$  PtMn(15)/ Co$_{70}$Fe$_{30}$(2.3)/ Ru(0.85)/ Co$_{40}$Fe$_{40}$B$_{20}$(2.4) is the synthetic antiferromagnet and FL $\equiv$ Co$_{20}$Fe$_{60}$B$_{20}$(1.8)  is the free layer.  Prior to patterning, the multilayers are annealed for 2 hours at 300\,$^\circ$C in a 1 Tesla in-plane magnetic field that sets the exchange bias direction for the SAF bottom layer parallel to the long (easy) axis of the nanopillar. 

The CoFeB/MgO based multilayer film composition is Si/ SiOx/ Ta(5\,nm)/ Co$_{20}$Fe$_{60}$B$_{20}(d)$/ MgO(1.1\,nm) /Ta(1\,nm)/ Ru(2\,nm), where the CoFeB layer thickness $d$ ranges from 0.9\,nm to 2.5\,nm. In order to minimize the sample-to-sample variations, the films were grown in a single run without changing the deposition parameters. The films were post-deposition annealed at 300\,$^\circ$C for 30\,minutes. We stress that the multilayers used in the MTJs and the CoFeB/MgO films are not only compositionally different but were also deposited in different sputter deposition systems. Nevertheless, the features of the temperature-dependent transitions described below are similar in the two systems, which suggests that the observed phenomena could be widespread in CoFeB/MgO based systems.

Fig.\,1(a) shows the hysteresis loops of the MTJ resistance versus magnetic field $B$ applied in the plane of the sample parallel to the long (easy) axis of the ellipse measured at $T$ = 4.2 K and $T$ = 295 K. The hysteresis loop shapes are consistent with both the FL and the SAF magnetic moments lying in the sample plane as illustrated in Fig.\,1(b). The TMR is 77\,\% at 295\,K and 97\,\% at 4.2\,K \cite{Parkin2004}. Fig.\,1(c) shows the temperature dependence of the MTJ resistance measured in the parallel state of the MTJ at $B=-33$\,mT. The MTJ resistance decreases with increasing temperature and exhibits a step-like onset at $T$ = 160\,K, that fades out at $T$ = 280\,K. We refer to this asymmetric feature as 'peak' for simplicity in what follows. Since we confirmed that the MTJ remains in the parallel state through the measurement temperature range, the resistance peak must be attributed either to modification of the layers' resistance or to a change in the interlayer tunneling probability. Whether the sharp transition near $T$ = 160\,K is of magnetic origin cannot be determined from electrical transport measurements alone, and magnetometry measurements are required. The low magnetic volume of the nanoscale MTJ does not allow for direct magnetometry measurements, and thus we employ the large-area CoFeB/MgO based film samples for temperature dependent magnetometry.

\begin{figure}[pt]
\includegraphics[width=1.0\columnwidth]{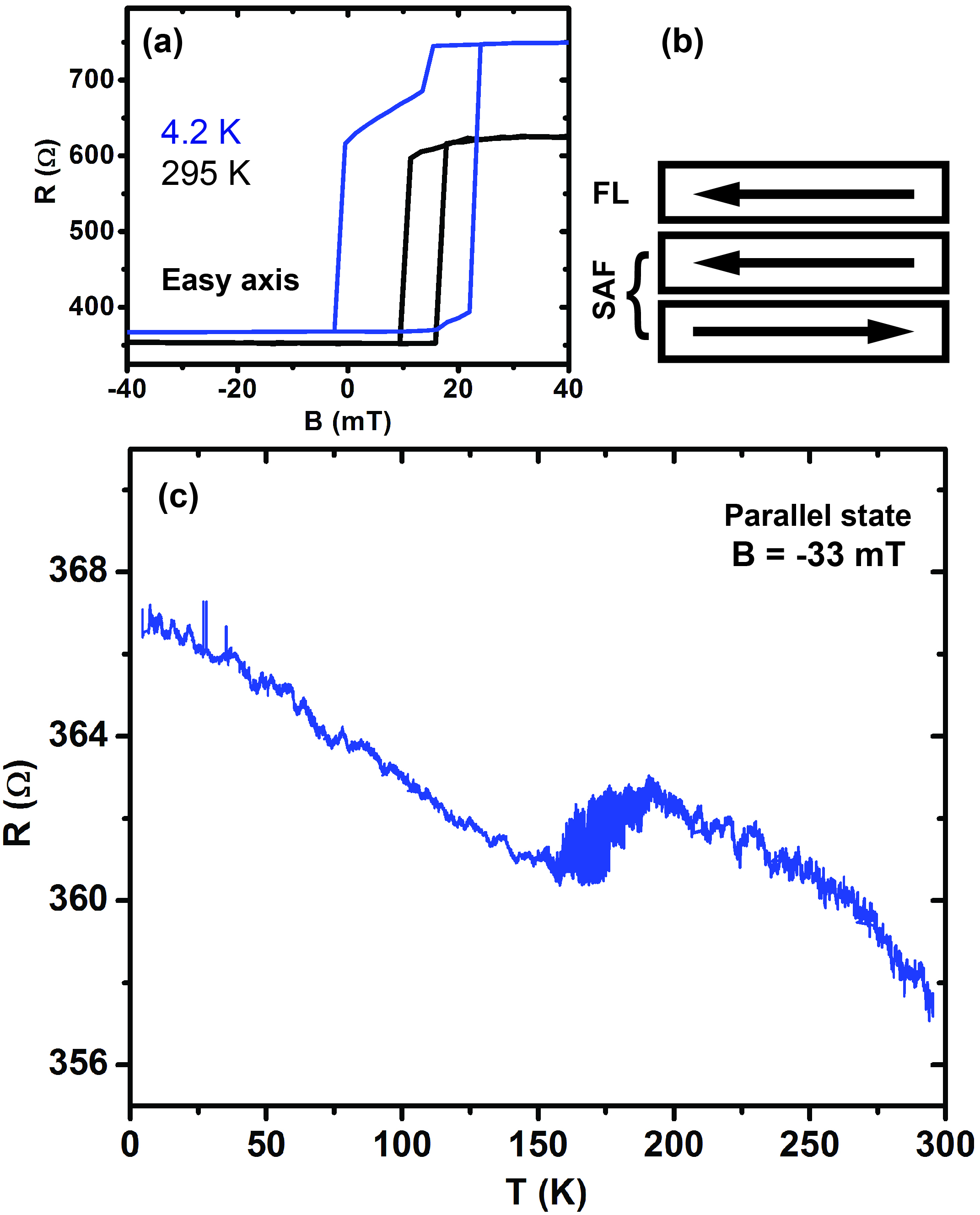}
\caption{\label{MTJ} (a)~MTJ resistance as a function of magnetic field applied parallel to the easy axis measured at $T$\,=\,4.2\,K and $T$\,=\,295\,K. At higher negative fields, the MTJ is in the parallel state depicted in~(b). (c)~ MTJ resistance versus temperature in the parallel state.}
\end{figure}

We use a superconducting quantum interference device (SQUID) to measure magnetization $M$ of the CoFeB/MgO films as a function of temperature in a saturating magnetic field applied in the film plane. Fig.\,2(a) shows magnetization of the sample with 1.3\,nm thick CoFeB layer normalized to its 5\,K value. The $M(T)$ data reveal a broad asymmetric peak with a precipitous step-like drop of $M$ at approximately 160\,K. We observe this broad peak in all  CoFeB/MgO films we studied (the CoFeB layer thickness ranging from 0.9\,nm to 2.5\,nm). The position and shape of the peak in $M(T)$ (Fig.\,2) are very similar to those of the $R(T)$ peak observed in the MTJ sample (Fig.\,1(c)). Since the measurements are performed in a saturating magnetic field, we conclude that the peak in $M(T)$ cannot be related to any magnetization switching process. Therefore, the peak must be attributed to the presence of an additional magnetic phase in the films (e.g. magnetic alloy or oxide). The peak's shape suggests that a magnetic phase with non-zero magnetization and critical temperature of approximately 280\,K is present in the sample. With decreasing temperature, this phase undergoes a first order transition to a phase with zero net magnetization (e.g. antiferromagnetic phase) at a temperature of approximately 160\,K. 

\begin{figure}[pb]
\includegraphics[width=1.0\columnwidth]{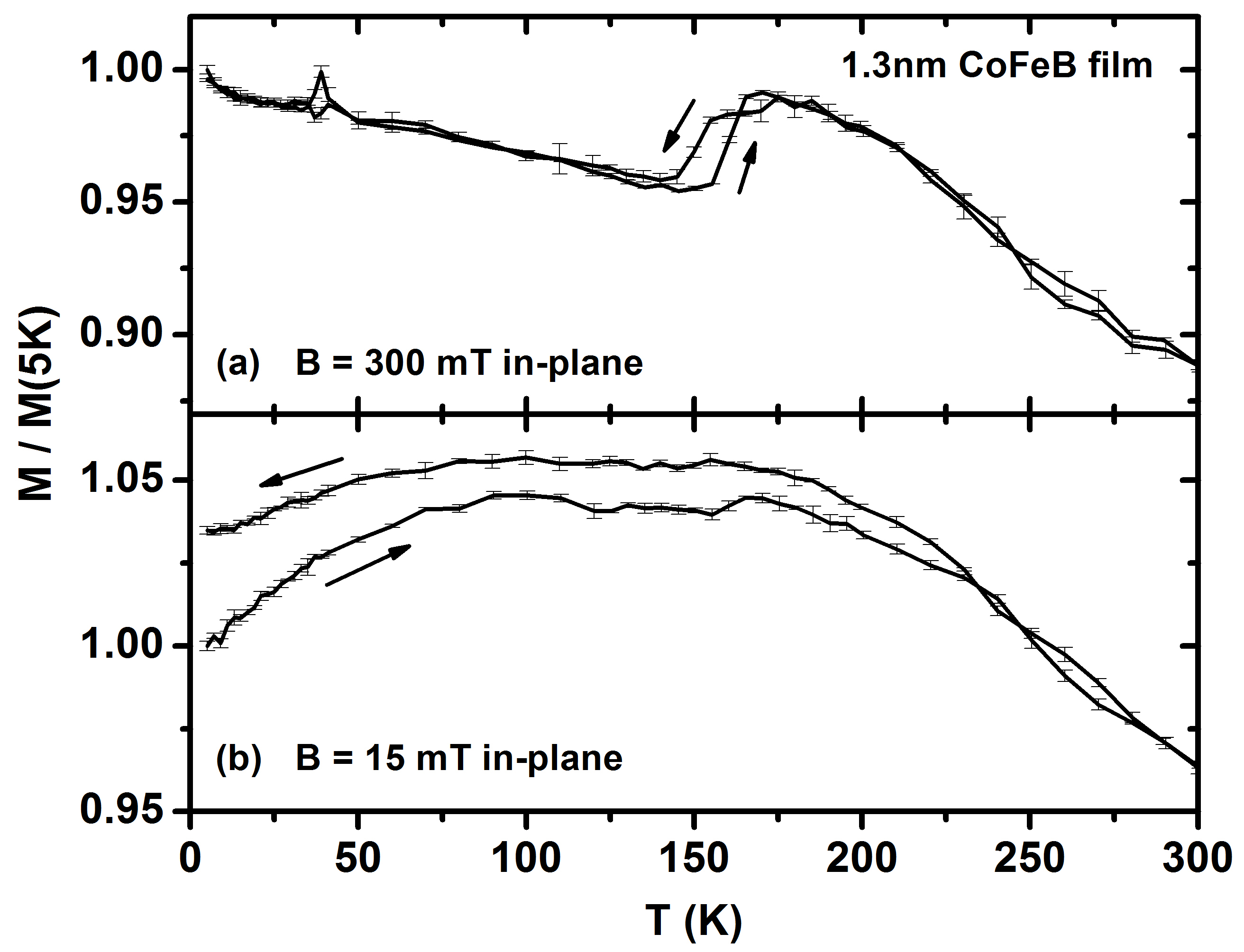}
\caption{Normalized magnetic moment of the 1.3\,nm thick CoFeB film as a function of temperature. After cooling down the sample in zero-field, the warm-up curve and consecutively the cool-down curve as indicated by the arrows are measured in the magnetic field applied in-plane.}
\end{figure}

To gain further insight into the magnetic phase transition in CoFeB/MgO films, we carry out electrical transport measurements with current and magnetic field applied in the film plane. The film resistance as a function of temperature is shown in Fig.\,3 for the sample with 1.1\,nm thick CoFeB layer. The resistance decreases with increasing temperature and exhibits relatively sharp transitions at temperatures near 160\,K. The magnitude of the observed resistance jumps ($\sim$4\%) significantly exceeds anisotropic magnetoresistance ($\sim$0.035\%) of the sample at 160\,K, and thus the observed resistance jumps cannot be explained by any CoFeB layer magnetization switching.  The resistance jumps must be attributed to intrinsic resistivity variations of the film. We also find resistance transitions for other CoFeB/MgO films with the CoFeB layer thicknesses between 0.9\,nm and 2.5\,nm. In Fig.\,3, three data sets are shown: The first $R(T)$ curve is measured during the cooling process. The resistance increases with two distinct jumps slightly below and above  160\,K. The second curve is taken during the sample heating process. With increasing temperature, the resistance decreases with two smoother transitions. The resistance jumps suggest that a fraction of the sample undergoes a first-order metal-insulator transition. In the subsequent temperature cycling, the resistance always shows jumps with some variation of the exact magnitude and position of the jumps from one temperature sweep to another. We have confirmed the presence of these unusual resistance jumps in  all CoFeB samples we studied via both 4-point and 2-point resistance measurements in a setup with verifiably low and stable contact resistances. We have also verified that the jumps are specific to the CoFeB/MgO system via measurements of other types of ultrathin metallic films such as Pt in the same measurement setup and found that such films exhibit the expectedly smooth non-hysteretic temperature dependence of resistance. The strong sweep-to-sweep variation of the resistance jump temperatures in the CoFeB samples point to a complex collective behavior possibly arising from interaction among coupled nanoscale magnetic oxide grains undergoing metal-insulator transition. Further studies are needed for detailed understanding of this complex behavior.

\begin{figure}[pt]
\includegraphics[width=1.0\columnwidth]{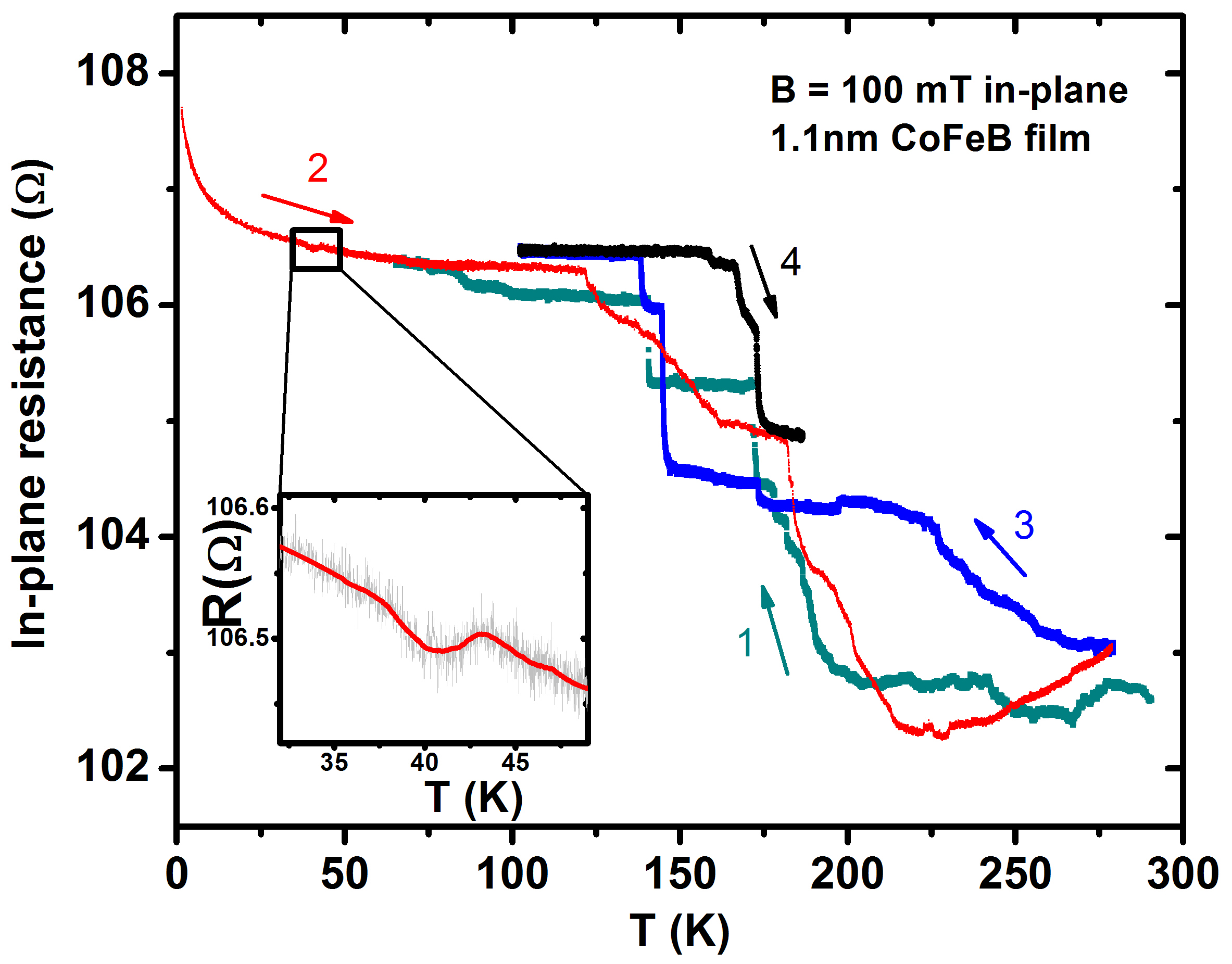}
\caption{\label{RT} Resistance of the 1.1\,nm thick CoFeB film. The arrows represent the temperature sweep direction. The numbers indicate the thermal cycle sequence.}
\end{figure} 

In Fig.\,3, a small feature in the resistance curve is observed at 40\,K. A similar small feature also appears  at 40\,K in the temperature dependence of the magnetization shown in Fig.\,2(a) for the 1.3\,nm CoFeB film. It is more pronounced at higher in-plane magnetic fields. A distinct peak is seen for in-plane external field of 300\,mT. For a lower field value of 15\,mT, it is much less prominent (Fig.\,2(b)). We attribute this peak to the presence of Co$_3$O$_4$ phase. Its enhanced magnetic susceptibility at the N\'{e}el temperature $T_N=40\,K$ can give rise to the field-dependent peak. It must be noted that $T_N$ of nanoscale Co$_3$O$_4$ would normally be expected to decrease. However, this phase is partially incorporated in a ferromagnetic matrix, which can also influence the N\'{e}el temperature. While this makes estimation of $T_N$ problematic, the potential presence of the cobalt oxide phase in the CoFeB/MgO films gives additional support to the picture of partial oxidation of the CoFeB layer \cite{Jen2006} and requires a discussion of possible oxide types that could form in the sample. 

Iron oxides are known to exhibit complex structural, magnetic, and electronic properties. Magnetite (Fe$_3$O$_4$), for instance, undergoes a phase transition at $T_{V}\sim$120\,K, known as Verwey transition \cite{IRM214, Verwey, Walz2002}. At temperatures above $T_{V}$, the ferrimagnetic magnetite possesses inverse spinel structure and electrical conductivity dominated by electron hopping between the mixed-valence octahedral sites. With decreasing temperature, the structure changes from cubic to monoclinic associated with an abrupt decrease of conductivity, lower magnetic susceptibility, and increased magnetocrystalline anisotropy while its magnetic anisotropy easy axis changes from $<$111$>$ to $<$100$>$ \cite{IRM214, Garcia}. Cation substitution (in particular by Co), oxygen vacancies, and stress are known to influence the transition temperature \cite{Abelian}. 
Another type of phase transition, known as Morin transition \cite{Morin}, is found in hematite (Fe$_2$O$_3$) at $T_M\sim 250$\,K. Above $T_M$, hematite is a weak ferromagnet \cite{Moriya, Dzya1958} with spins lying in the basal $c$-plane of the rhombohedral lattice \cite{IRM201,Moriya, Dzya1958}. With decreasing temperature, the spin directions flip to the c-axis, and hematite becomes  an antiferromagnet \cite{IRM201, Park}. Its conductivity \cite{Nakau1960} is expected to drop \cite{Morin, Anderson1959}, however this effect can be altered due to the strong dependence of the conductivity on impurities, dopants, and stoichiometric defects in the hematite \cite{Acket1966, Gardner1963}. Besides stress \cite{Wayne,Schroeer1967, Bengoa}, inhomogeneities, and oxygen vacancies \cite{Ma2012}, the transition temperature  depends strongly on the crystallographic grain size \cite{Yamamoto, IRM201,Ozdemir2008}. Reduction of $T_M$ to values of $\sim$160\,K occurs in small crystallites of a few nanometers in diameter \cite{Ozdemir2008,Park}, which is likely to be the grain size of the magnetic oxide in our samples.

The properties of these magnetic phase transitions in Fe oxides are generally consistent with the phase transitions seen in our magnetometry and electric transport measurements on CoFeB films. Therefore, our results strongly suggest that a thin oxide layer is formed at the CoFeB/MgO interface. This oxide layer exhibits a phase transition from an insulating antiferromagnet at low temperatures to a conductive weak ferromagnet at higher temperatures with the transition temperature near 160\,K. The transition is more prominent at higher fields (Fig.\,2) and exhibits a temperature hysteresis (Fig.\,3) \cite{Ozdemir2008}. These observations are consistent with the Morin transition in nanoscale hematite grains.

Since we observe the magnetic phase transitions in two different CoFeB-based types of samples (nanoscale MTJs and thin films), it is likely that the interfacial oxide formation is generally present in annealed CoFeB/MgO systems \cite{Bengoa, Chen2014}. The presence of an interfacial magnetic oxide layer is also consistent with our recent studies of CoFeB/MgO films by room-temperature ferromagnetic resonance \cite{BarsukovPMA}. These studies demonstrated that the effective PMA of a CoFeB layer depends on external magnetic field. This unusual field-dependent PMA was attributed to exchange coupling of the CoFeB magnetization to a magnetic phase with a high saturation field. Such a phase is consistent with a magnetic oxide phase with a critical temperature near room temperature \cite{BarsukovPMA, Shaw}. 

In conclusion, magnetometry and electrical transport measurements of ultrathin films and magnetic tunnel junctions based on Ta/CoFeB/MgO multilayers reveal magnetic phase transitions below room temperature. Our observations suggest that a fraction of the multilayer undergoes a phase transition from a conductive weakly ferromagnetic state at higher temperatures to an insulating antiferromagnetic state at lower temperatures with the transition temperature near $\sim$160\,K. This transition is consistent with the Morin transition in nanoscale hematite grains, which suggests that an iron-based magnetic oxide layer is formed at the CoFeB/MgO interface. We stress that the multilayers studied here were prepared under typical conditions used in fabrication of MTJs for STT-RAM memory elements -- they were deposited by magnetron sputtering and annealed at 300\,$^\circ$C. We, therefore, expect the interfacial magnetic oxides to be generally present in CoFeB-based STT-RAM elements. The existence of this magnetic oxide layer should be taken into account in quantifying  such important properties as effective saturation magnetization \cite{Liu,Jen2006}, effective PMA \cite{BarsukovPMA, Tsai,Yamanouchi}, Gilbert damping \cite{Devolder, BarsukovIEEE, Liu,ShawPMAalpha}, and the thickness of magnetically dead layers \cite{Devolder,Yamanouchi,Jen2006}.

We thank J. Langer for magnetic multilayer deposition. This work was supported in part by FAME, one of six centers of STARnet, a Semiconductor Research Corporation program sponsored by MARCO and DARPA, by Intel through grant No. 2011-IN-2152  as well as by NSF through Grants No. DMR-1210850 and No. ECCS-1002358. We acknowledge the Center for NanoFerroic Devices (CNFD) and the Nanoelectronics Research Initiative (NRI) for partial funding of this work. Funding by the DFG/NSF in the framework of the Materials World Network program is also acknowledged.  A.\,M.\,G. thanks CAPES Foundation, Ministry of Education of Brazil for financial support.

\end{document}